%% file: main.tex
\documentclass{astlb}

\usepackage{graphicx} 
\usepackage{amsmath,amssymb} 
\usepackage{epsfig,graphics} 
\usepackage{rotating} 
\usepackage{ulem}

\usepackage{epstopdf}
\usepackage{pdfpages}

\usepackage{longtable}
\usepackage{caption}
\usepackage{dcolumn}
\usepackage{pdflscape}

\usepackage{hyperref}

\usepackage{comment}
\usepackage{ulem}
\usepackage{color}
\usepackage{tcolorbox}
\usepackage{xspace}

\newcommand {\be}{\begin{equation}} 
\newcommand {\ee}{\end{equation}}

\newcommand{\Msun}{M_\odot}

\def\FX0.52{F_{{\rm X},0.5-2}}
\def\LX210{L_{{\rm X},2-10}}

\def\red#1{\textcolor{red}{#1}}
\def\БР#1{\red{\textbf{#1}}}

\newcommand{\zspec}{z_{\rm{spec}}}
\newcommand{\zph}{z_{\rm{ph}}}

\newcommand{\aosx}{\alpha_{\rm ox}}

\newcommand{\nuo}{\nu_{2500\,\textup{\AA}}}
\newcommand{\Lo}{L_{2500\,\textup{\AA}}}

\newcommand{\nusx}{\nu_{2\,{\rm keV}}}

\newcommand{\Lsx}{L_{2\,{\rm keV}}}

\newcommand{\Lnx}{L_{2-10\,{\rm keV}}}

\newcommand{\Mbh}{M_{\rm BH}}

\newcommand{\srg}{{SRG}\xspace}
\newcommand{\xmm}{{XMM-Newton}}
\newcommand{\chandra}{{Chandra}}
\newcommand{\rosat}{{ROSAT}}
\newcommand{\swift}{{Swift}}

\newcommand{\wise}{{WISE}}

\newcommand{\erosita}{{eROSITA}}
\newcommand{\srge}{{SRG/eROSITA}\xspace}

\newcommand{\BTA}{BTA\xspace}
\newcommand{\scorpioii}{{SCORPIO-2}\xspace}
\newcommand{\panstarrs}{Pan-STARRS}

\newcommand{\srgz}{SRGz}
\newcommand{\daleqo}{DaLeQo}

\def\kev2l{L_{2keV}}
\def\l2500A{L_{2500\mathring{A}}}

\newcommand{\fnufiveGHz}{f_{\nu,\,5\,GHz}}
\newcommand{\fnufothsndA}{f_{\nu,\,4400 \mbox{\small \AA}}}

\newcommand{\theqso}{SRGE\,J170245.3+130104\xspace}
\newcommand{\mqso}{CFHQS\,J142952+544717}
\newcommand{\mqsos}{SRGE\,J142952.1+544716}

\newcommand{\Ho}{$H_0$=69.6~km/s/Mpc\xspace}
\newcommand{\Om}{$\Omega_M$=0.286\xspace}

\begin{document} 

\journalinfo{2021}{47}{3}{155}[173]
\title{Discovery of the most X-ray luminous quasar \theqso at redshift \mbox{\small z}$\approx5.5$ }
\author{G.A.~Khorunzhev\email{horge@iki.rssi.ru}\address{1}, A.V.~Meshcheryakov\address{1,2}, P.S.~Medvedev\address{1}, V.D.~Borisov\address{1,3}, R.A.~Burenin\address{1}, R.A.~Krivonos\address{1}, R.I.~Uklein\address{4}, E.S.~Shablovinskaya\address{4}, V.L.~Afanasyev\address{4}, S.N.~Dodonov\address{4}, R.A.~Sunyaev\address{1,5},  S.Yu.~Sazonov\address{1}, M.R.~Gilfanov\address{1,5}
$^1$\it{Space Research Institute RAS, Moscow, Russia\\}
$^2$\it{Kazan Federal University, Kazan, Russia\\}
$^3$\it{Faculty of Computational Mathematics and Cybernetics, Lomonosov Moscow State University, Moscow, Russia\\}
$^4$\it{Special Astrophysical Observatory RAS, Nizhnij Arkhyz, Russia\\}
$^5$\it{Max Planck Institute for Astrophysics, Garching, Germany}
}
\shortauthor{Khorunzhev et al.}  
\shorttitle{Extremely bright quasar \theqso at $z=5.5$} 
\submitted{December 15, 2020}

\begin{abstract} 
\theqso \ was discovered by the \erosita\ telescope aboard the \srg\ space observatory on March 13-15, 2020 during the first half-year scan of its all-sky X-ray survey. The optical counterpart of the X-ray source was photometrically identified as a distant quasar candidate at $z\approx5.5$. Follow-up spectroscopic observations, done in August/September 2020 with the SCORPIO-II instrument at the BTA 6-m telescope, confirmed that \theqso\ is a distant quasar at redshift $\zspec =5.466\pm0.003$. The X-ray luminosity of the quasar during the first half-year scan of the eROSITA all-sky survey was $3.6^{+2.1}_{-1.5}\times 10^{46}$~erg/s (in the 2--10~keV energy range), whereas its X-ray spectrum could be described by a power law with a slope of $\Gamma=1.8^{+0.9}_{-0.8}$. Six months later (September 13--14, 2020), during the second half-year scan of the eROSITA all-sky survey, the quasar was detected again and its X-ray luminosity had decreased by a factor of 2 (at the $\approx 1.9\sigma$ confidence level). \theqso\ proves to be the most X-ray luminous among all known X-ray quasars at $z>5$. It is also one of the radio-loudest distant quasars (with radio-loudness $R\sim10^3$), which may imply that it is a blazar. In the Appendix, we present the list of all $z>5$ quasars detected in X-rays to date.

\keywords{quasars, X-rays, sky surveys, photo-z, spectroscopy, BTA, eROSITA.}
\end{abstract} 
\clearpage

\section{Introduction}
Launched July 13, 2019, the \srg\ X-ray observatory \citep{sunyaev20,predehl20} is orbiting around Lagrangian point L2 of the Earth-Sun system. The main purpose of the observatory is 
the four-year X-ray All-Sky Survey over a wide range of energies 0.2--30 keV. In the course of the sky survey it is expected to discover with the \srge\ telescope \citep{predehl20} 
about three million active galactic nuclei (AGNs), including distant quasars 
\citep{kolodzig13a, kolodzig13b}, about 100,000 galaxy clusters and groups, and few $10^5$~X-ray 
sources of various nature in our Milky Way Galaxy. 
After four year of operation, the \srg\ All-Sky Survey is expected to be about 25 times more sensitive in the soft X-ray band (0.5-2 keV) than the previous X-ray survey conducted by the \rosat\ \citep{trumper82} satellite in the early 1990s, 
and will help to solve a number of important problems in modern astrophysics and cosmology. One of the main aims of the SRG/eRosita survey is to search unique objects, 
whose spectral characteristics are outstanding among astronomical sources of their class. 
In particular, the detection and detailed study of extremely bright quasars will shed light on the growth history of the most massive black holes in the Universe.

In June 2020, the \srg observatory completed first half-year scan of X-ray All-Sky Survey. For 
all new X-ray sources discovered with the \erosita \ telescope, most probable optical 
counterparts were identified by \srgz \ machine learning system \citep{mesch21}. This software 
automatically searches and classifies the most probable optical counterparts of extragalactic 
X-ray sources, and estimates their photometric redshifts based on data from optical and 
infrared sky surveys. The \srgz \ system was created in the Science Working Group of RU eROSITA 
consortium on X-ray source detection, identification and eROSITA source catalog. 

The high performance of SRGz predictions was confirmed by a set of follow-up spectroscopic 
bservations of SRGz candidates: distant quasars at $z\sim4$ and quasars at lower redshifts 
$z\sim1\div3$ from deep X-ray survey of the Lockman Hole extragalactic field (observed by 
eROSITA during the Performance Verification phase) and from the first scan of eROSITA All-Sky 
Survey. The optical spectra of the candidates were obtained on AZT-33IK 1.6-m telescope of the 
Sayan Observatory \citep{khorunzhev20}, Russian-Turkish 1.5-m telescope (RTT-150, \citealt{bikmaev20}) and 2.5-m KGO telescope of the Lomonosov Moscow State University \citep{dodin2020}.

Of particular interest are bright in X-rays 
(with luminosity $\Lnx>5\times 10^{45}$ erg/s in the 2--10~keV energy range) and distant ($z>3$) quasars, which are rarely found in deep X-ray surveys of small area and were not be detected 
in previous X-ray All-Sky Survey conducted by \rosat \ observatory 
(because of its insufficient depth). Already at the very beginning scans of the X-rya survey, \srg\ observatory managed to find a unique distant quasar \mqso\ = \mqsos, 
which happened to be the brightest in X-rays among known quasars at $z>6$ \citep{medvedev2020}.

The  Science  Working  Group  of  RU eROSITA consortium for Active Galactic Nuclei Studies has 
created a separate observational program called\footnote{Name 'daleqo' was invented by 
V.D.Borisov, from Russian it translates as 'far away'} \daleqo \ (Distant and Luminous eROSITA 
Quasi-stellar objects, \citealp{khorunzhev21}) dedicated to spectroscopy of SRGz candidates for 
distant ($z>3$) and bright ($L_X\gtrsim5\times 10^{45}$\,erg/s) X-ray quasars. This paper 
describes the discovery of the most outstanding (so far) object found in the \daleqo program --- the brightest X-ray and Radio distant quasar \theqso at $z>5$, identified by spectral 
observations on the 6th BTA optical telescope.

In the paper the following cosmological parameters were adopted (for calculation of luminosities): \Ho \Om \ \citep{bennett14}.

\section{X-ray data}
The \theqso source was discovered by eROSITA telescope during the first half-year scan of the X-ray All-Sky Survey, the position of the source was scanned nine times between March 13 and 15, 2020. The total time the source was scanned 
with the \erosita{} telescope was 256\,s. The source was found in the energy range 0.3--2.2 keV at coordinates RA$=255.688844$ deg, DEC$=13.017685$ deg with localization accuracy of $7$~arcsec (95\%). In the  0.3--2.2 keV energy range 30 counts were recorded inside a 60 arcsec radius circle (centered on the X-ray source position), with an expected background counts of 7.4. Approximation of the counts distribution in the source vicinity by the point spread function model gives a statistical significance of source detection $8.5\sigma$.

\begin{figure*}
\centering
\includegraphics[width=1.0\textwidth]{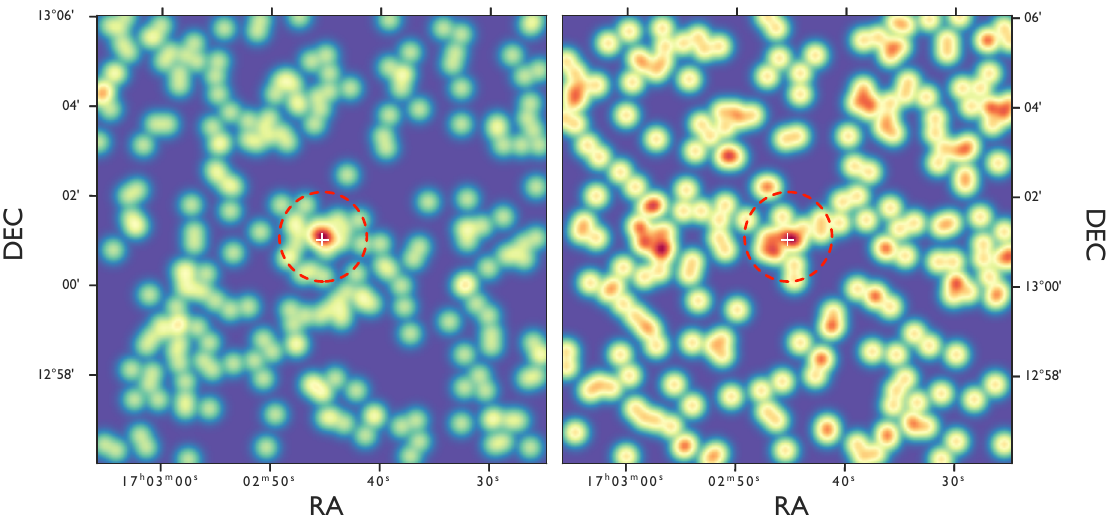}
\caption{X-ray images of the \srge region $10^{\prime} \times 10^{\prime}$ in the energy range 0.3--2.2 keV, 
centered on the optical source coordinates \theqso (white cross). Left panel --- image from the first half-year scan of \srg/\erosita{} All-Sky Survey, right panel --- image from the second half-year scan. Aperture with a radius of $1^{\prime}$ is shown by red circle. Images are smoothed with a Gaussian filter ($\sigma=8^{\prime\prime}$).}
\label{fig:erosita_img}
\end{figure*}

In the second half-year scan of the X-ray All-Sky Survey, the position of the source was observed by the \srg observatory from September 13 to 14, 2020. The total exposure time was 268\,s, during which only 18 counts in the 0.3--2.2 keV energy range were obtained within 60 arcsec aperture. 18 counts with 7.3 expected background counts give detection significance $3.7\sigma$ (see Figure \ref{fig:erosita_img}). The position of the source from the second scan appeared to be consistent (within uncertainty) with the coordinates measured in the first scan.

The primary processing of data from the \erosita{} telescope was carried out by software developed at IKI RAS by using components of the eSASS system (Max Planck Institute for Extraterrestrial Physics, Germany). The X-ray photon map of the $10'\times10'$ region centered at the source is shown in Fig.~\ref{fig:erosita_img}. The source spectrum was 
extracted from a circular aperture with a 35 arcsec radius. Background spectrum was estimated in a ring centered at the source with inner and outer radii of 85 and 435 arcsec, respectively. Other sources detected in 1+2 scans data and falling inside the background region were masked out using a circular aperture with a radius of 40 arcsec. 
The obtained spectra were approximated by using the standard tools of the {\sc XSPEC} \citep[version 12.11,][]{Arnaud1996} software package, using C-statistics \citep{Cash1979}  
modified for data with a Poisson background (the so-called W-statistics, see the {\sc XSPEC} documentation for details\footnote{\url{https://heasarc.gsfc.nasa.gov/xanadu/xspec/manual}}). 
The energy channels of the spectra were grouped so that the number of counts in each channel was at least one. This procedure was performed using a standard {\sc grouppha}\footnote{\url{https://heasarc.gsfc.nasa.gov/ftools}} tool.

\begin{table*}
\centering
\caption{Parameters of the best approximation of the \theqso X-ray spectrum according to the first scan of \srg/\erosita{} All-Sky Survey. The errors are given at 90\% significance level. In the "goodness" column,  the probability of obtaining $C$-stat less than the best $C$-stat value for a given model due to statistical fluctuations is shown. This probability characterizes the quality of the model's description of the data; for an adequate model one should expect a "goodness" value of $\sim 50\%$. The Akaike Information Criterion (AIC) is calculated using the formula $2n - 2\ln(L)$, where $n$ is the number of free parameters of the model, and $\ln(L)$ is the likelihood function logarithm; $- 2\ln(L)$ is equal to the value of the $C$-stat (according to its definition in XSPEC package). The model with a smaller AIC value is more preferable. }
\label{tab:bf}
\begin{tabular}{l c  c  c  c  c r}
\hline
{\bf Model}  &	{\bf Parameter} & {\bf Value} &	{\bf C-stat}/{\bf d.o.f} & {\bf goodness} & {\bf AIC} \\
\hline
\\
 & $N_H^{*}$ &	$5.1 \times 10^{20}$ cm$^{-2}$ & &  &  \\
 {\textsc{tbabs*cflux*pow}} &  $F^{cflux}_{05-2}$ & $1.03_{-0.4}^{+0.51} \times 10^{-13}$ erg cm$^{-2}$ sec$^{-1}$ & 23.18/17 & 53.9\% &  27.18  \\
 &  $\Gamma$  & $1.79_{-0.81}^{+0.86}$  & & \\
     
\\
\hline
\\
 & $N_H^{*}$ & $5.1 \times 10^{20}$ cm$^{-2}$ & & \\
 & $F^{cflux}_{05-2}$ & $1.12_{-0.44}^{+0.57} \times 10^{-13}$ erg cm$^{-2}$ sec$^{-1}$ &  & \\
{\textsc{tbabs*cflux*zphabs*pow}} & $N_H^{z}$ & 

$83.8_{-60.1}^{+119.5} \times 10^{22}$ cm$^{-2}$ & 14.95/16 & 8.8\% & 20.95 \\
 & $\Gamma$ & $5.46_{-2.61}^{+4.56}$ & & \\
 
\\
\hline
\\
 & $N_H^{*}$ & $5.1 \times 10^{20}$ cm$^{-2}$ & &\\
 & $F^{cflux}_{05-2}$ & $0.99_{-0.38}^{+0.50}\times 10^{-13}$ erg cm$^{-2}$ sec$^{-1}$ & & \\
     & $\tau_{\mathrm{edge}}$ & $<8$ &  \\
{\textsc{tbabs*cflux*(zedge*pow }} & $\Gamma$ & $1.45_{-0.82}^{+0.97}$ & 19.98/15 & 51.8\% & 27.98 \\
 {\textsc{ + zgauss)}}    & $EW_{\mathrm{Fe\,K}\alpha}$ & $<$670 eV & & & \\
     & $\sigma_{\mathrm{Fe\,K}\alpha}^{*}$ & 10 eV & & &\\
     
\\
\hline
\end{tabular}\\
$^{*}$  The parameter is fixed.
\end{table*}

\begin{table*}
\onecolumn
\begin{longtable}{cD{.}{.}{6}D{.}{.}{6}cccccc}
\caption{Optical companion properties and photometric redshift prediction} 
\label{tab:optsample} \\
\hline
\hline

\multicolumn{1}{c}{OBJID SDSS} &
\multicolumn{1}{c}{RAopt} &
\multicolumn{1}{c}{DECopt} &
\multicolumn{1}{c}{sep} &
\multicolumn{1}{c}{$i^\prime_{\rm psf}$} &
\multicolumn{1}{c}{$C_{\rm ph}$} &
\multicolumn{1}{c}{$z_{\rm ph}$} &
\multicolumn{1}{c}{$\rm zConf$} \\
\hline
\endhead
\endfoot

1237665106509826046&255.688797& +  13.017288& 1.4 &22.04&QSO&5.486&0.84\\ 
                
\hline
\end{longtable}
\caption*{{\bf Note.} OBJID SDSS, RAopt and DECopt --- unique
identification number and equatorial coordinates of the optical source from the SDSS DR14 photometric catalog, sep --- angular distance between X-ray and Optical sources (arcsec), $i^\prime_{psf}$ --- apparent magnitude in the $i^\prime$ SDSS filter; SRGz predictions:
$C_{ph}$ --- photometric classification (star, galaxy, quasar), $\zph$ --- photometric redshift, $zConf$ --- reliability of the photo-z measurement.}
\twocolumn
\end{table*}

Spectral analysis of the data was performed in the energy range of 0.3--5 keV. Only data from the first half-year scan were used to fit the X-ray spectrum, since the number of counts in the second scan is not sufficient for a meaningful spectral analysis. A joint analysis of the data from the two 1+2 scans was not performed due to the possible variability of the source spectrum. 

\begin{figure}
\centering
\includegraphics[width=1.0\columnwidth]{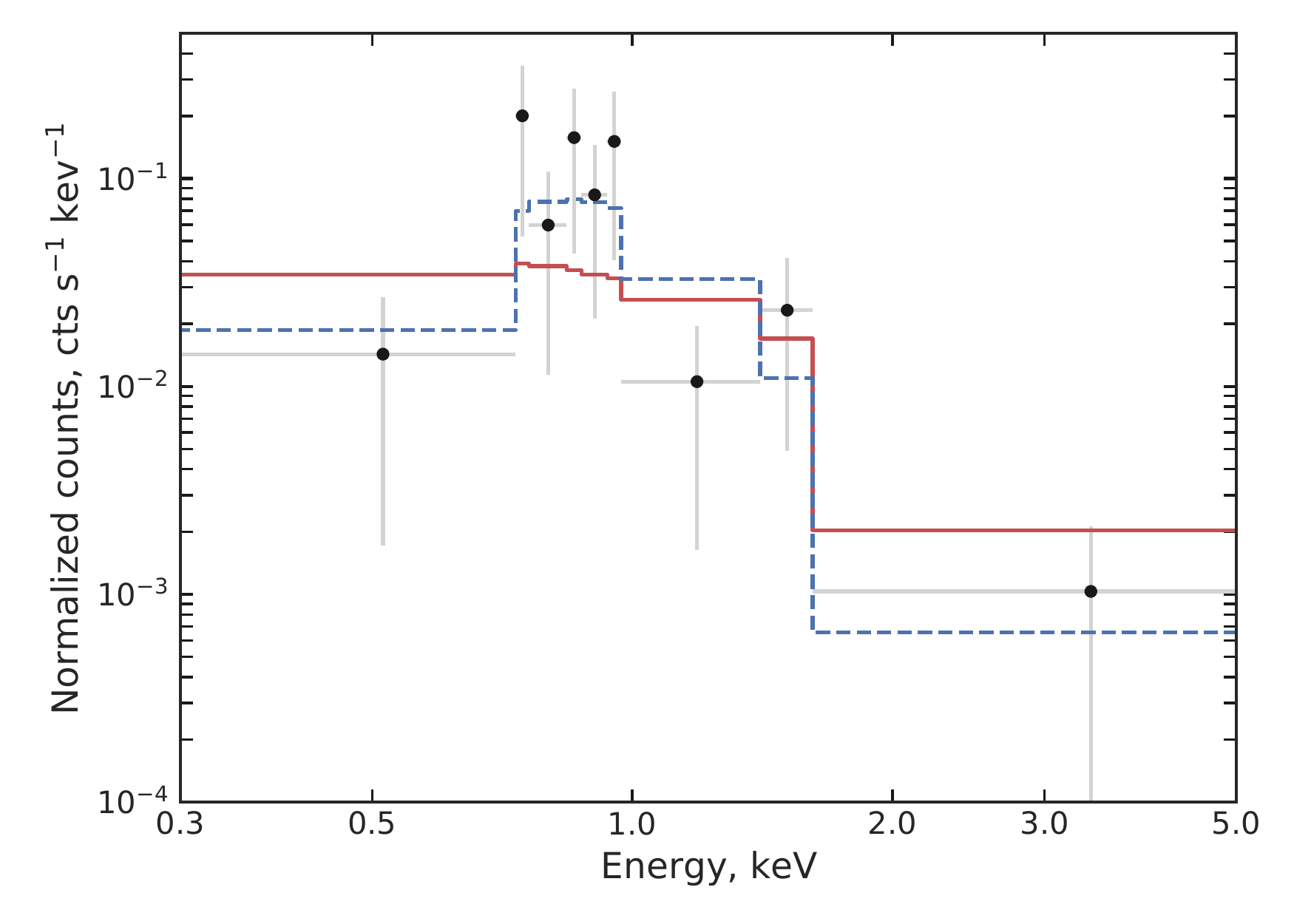}
\caption{X-ray spectrum of the \theqso quasar from the first half-year scan of the \srge{} All-Sky Survey. 
Red solid line shows the best fit power law model with Galactic absorption, 
blue dotted line --- model with additional absorption in the quasar reference frame (Table \ref{tab:bf}).
}
\label{fig:erosita_spec}
\end{figure}

We used simple absorbed power law spectral model to fit data. The hydrogen column density $N_{\rm H}$ was fixed to the Galactic absorption value according to HI4PI \citep{HI4PI_collab} maps: $N_{\rm H}=5.10\times10^{20}$~cm$^{-2}$. The derived parameters of the best fit model are given in Table ~\ref{tab:bf}, model spectrum is shown in Figure ~\ref{fig:erosita_spec} with the red line. As can be seen from Table \ref{tab:bf}, derived ''goodness of the fit'' value indicates that the proposed model adequately describes the data. 

Relatively small number of source counts recorded from \theqso{} does not allow one to investigate in detail a more complex spectral models. Nevertheless, we applied two more models to the data, taking into account possible internal absorption and spectral features associated with reflected component in the quasar spectrum.

Firstly, we added multiplicative intrinsic absorption component (in the quasar rest frame $z=5.466$ with a free absorption parameter) to the basic absorbed power law spectral model. Such a modification of the model (shown as blue dashed line in Fig.~\ref{fig:erosita_spec}) leads to a noticeable decrease (as compared to the basic model) both in the value of $C$-statistics ($C$-stat$ = 14.95$ at d.o.f.$ = 16$) and in the information criterion value $AIC$ \citep{AIC}. However, such a model requires large and apparently unrealistic absorption and photon spectral index values: $N_{\rm H} \approx 8\times10^{23}$~cm$^{-2}$, $\Gamma \approx 5$, although determined with large statistical errors. Therefore, our analysis does not allow us to make an unambiguous conclusion about the presence of internal absorption in the source.

We studied possible contributions to the source spectrum from reflected emission --- from a neutral optically thick medium (e.g., accretion disk or molecular torus). We added to the basic power law  model the Fe fluorescence line at an energy of 6.4 keV and Fe K-edge absorption at an energy of 7.1 keV (in the quasar rest frame). The resulting best fit model formally describes  data somewhat better ($\delta C$-stat$=-3.2$) than a simple power law  model, but from small difference in AIC (see Table ~\ref{tab:bf}) one can conclude that data does not provide sufficient information to justify this complication of the model. Thus, more accurate conclusions regarding the shape or presence of any features in the spectrum \theqso require X-ray data with a higher signal-to-noise ratio. 

In order to calculate internal (unabsorbed) X-ray luminosity of \theqso we used a basic power law model described above, i.e., we corrected for absorption in the Galaxy's interstellar medium.
We obtained unabsorbed X-ray luminosity of the source  $3.6_{-1.5}^{+2.1} \times 10^{46}$ erg/c in the 2--10 keV energy range (in the quasar rest frame).

In the second half-year scan of \erosita{} All-Sky Survey source counts were decreased by a factor of 2. More precisely, the 0.3--2.2 keV countrate drop 
by a factor of $\approx 2.3$ (with statistical confidence $\approx 1.9\sigma$). By using the basic spectral model shape from the first \erosita{} scan data fit, one can estimate X-ray flux and luminosity (in the 2--10 keV energy range) in the second \erosita{} scan as $5.0^{+4.5}_{-3.1}\times 10^{-14}$ erg/s/cm$^2$ and $1.7^{+1.5}_{-1.1}\times 10^{46}$ erg/s$^2$, respectively.

\section{Selection of distant X-ray quasars candidates}

We made a cross-match of \srge\ sources from the first half-year X-ray scan at $R_{\rm match}=$10 arcsec radius (corresponds to median 98\% localization error of X-ray sources) with photometric catalogs from three optical surveys: DESI Legacy Imaging Survey DR8 (DESI LIS, \citealt{dey19}), PanSTARRS1 DR2 \citep{chambers16}, SDSS DR14 \citep{abolfathi18}. We used \wise{} \citep{lang16} forced infrared photometry measurements from the DESI LIS catalog.

Optical data from different optical catalogs were combined into a master photometric catalog by performing a cross-match of all optical sources within a 1'' radius, photometric measurements with the highest signal-to-noise ratio were selected separately in each filter. Further, the \erosita{} X-ray sources with more than one optical object in DESI LIS (within the selected $R_{\rm match}$ radius) were excluded from consideration.

The resulting list of optical candidates was processed by the \srgz{} system, which operates over the entire Eastern Galactic hemisphere region of the \erosita\ X-ray survey and automatically analyzes photometry and positions of optical objects in the X-ray sources vicinity. 
\srgz\ use tree-based machine learning algorithms (Gradient Boosting and Random Forest, see \citealp{mesch18}), which are trained on samples of quasars, galaxies, and stars from the 
SDSS spectroscopic catalog, sample of distant $z>5$ quasars \citep{ross20} and sample of GAIA DR2 stars  associated with 3XMM DR8 sources. For more details on the principles of operation of \srgz\ and the algorithms implemented in it see \cite{mesch21}.

Based on the \srgz{} predictions, we selected optical candidates with the photometric class "quasar"\ and
photometric redshift $z_{\rm ph}>5$ measured with high confidence $zConf$ (calculated for each object from the full conditional probability distribution $p(z|x)$ as the integral in the neighborhood of photo-z point estimate~$\zph\pm0.06(1+\zph)$). 

\begin{figure}
    \centering
    \hbox{}
    \includegraphics[width=0.9\linewidth]{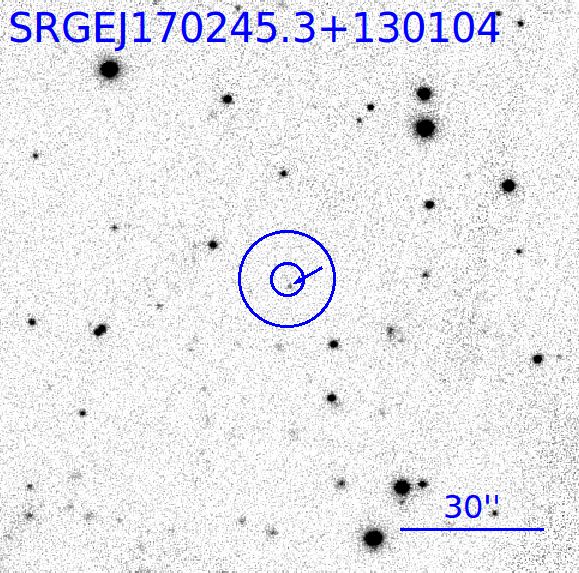}
    \caption{
    The 2'$\times$2' image in the $i_{PS}$ Pan-STARRS filter. The arrow indicates the position of optical companion \theqso. The radius of the small circle corresponds to the $1\sigma$ localization region. The radius of the large circle (10~arcsec) determines the size of the region where optical companion of X-ray source was searched. 
    }
    \label{fig:fchart}
\end{figure}

The X-ray source \theqso and its associated optical companion (see ~Table~\ref{tab:optsample}) were selected by SRGz system as the most reliable candidate for a distant X-ray quasar at $z>5$. \theqso{} has single optical source within 95\% X-ray localization radius, located 1''.5 from X-ray source coordinates. At Figure ~\ref{fig:fchart} we show $i_{PS}$-band archival image of \theqso{} region from Pan-STARRS survey. 

\begin{figure*}
    \centering
    \hbox{}
    \includegraphics[width=1.0\linewidth]{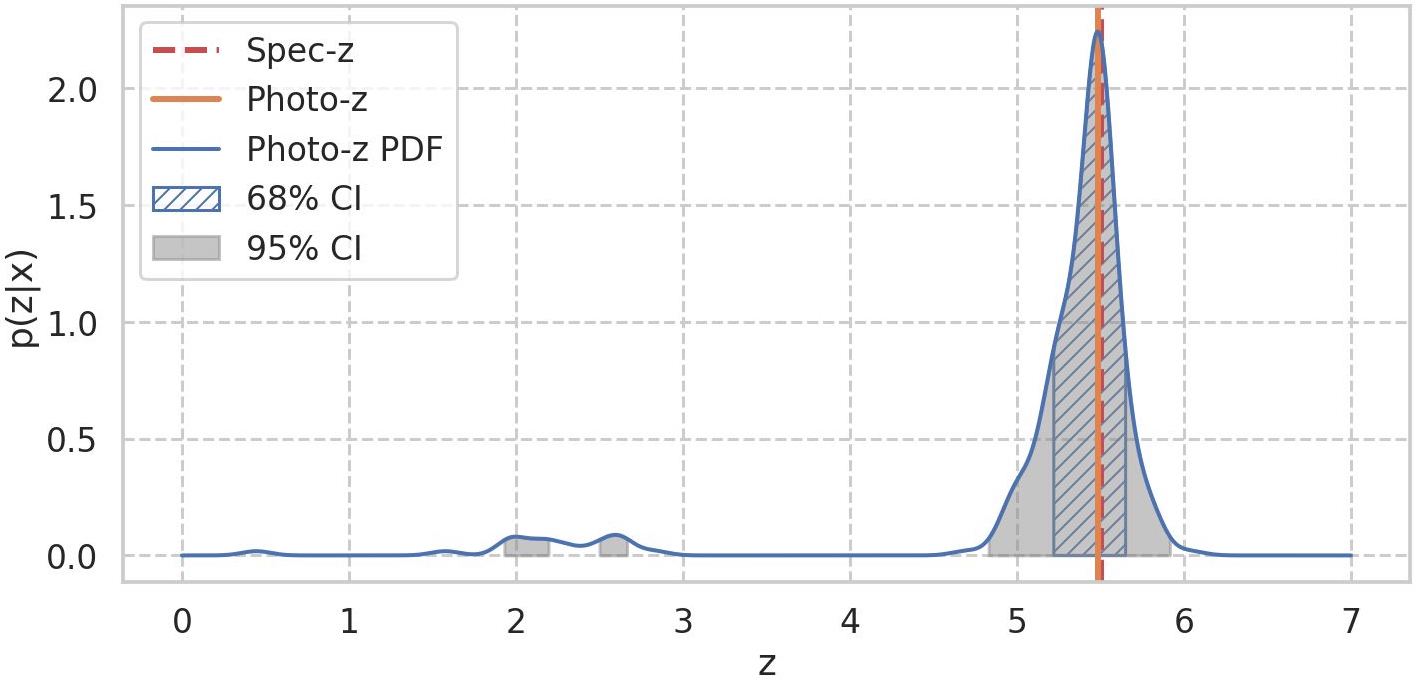}
    \caption{Full probabilistic photo-z prediction $p(z|x)$ obtained by the SRGz system for optical companion of X-ray source \theqso. The photo-z point estimate $z_{\rm ph}$ --- solid vertical line); shaded areas show confidence intervals with 68\% and 95\% significance levels; spectral redshift measured at the BTA telescope  --- dashed vertical line.}
    \label{fig:srgz_pdf}
\end{figure*}

At Figure ~\ref{fig:srgz_pdf} we show a full probability density function $p(z|x)$ predicted for optical companion of X-ray source \theqso, confidence intervals with 68\% and 95\% significance levels (shaded areas), point photo-z prediction (solid vertical line), and the spectral redshift measured at the BTA telescope (dashed vertical line). The conditional probability density function $p(z|x)$ shows a sharp peak at $\zph=5.486$. The spectral measurement of quasar redshift (see below) agrees perfectly with the photometric estimate obtained by the SRGz system.

Note that the optical companion of \theqso has not  been identified previously in the literature as a quasar candidate. The optical source has a radio companion in the NVSS survey at ~1.4~GHz. \citep{condon98}.

\section{\BTA observations}

The distant X-ray quasar photometric candidate was included in the observational program of the 6-m telescope \BTA. 

Observations at \BTA were conducted with the  \scorpioii spectrograph \citep{afanasev11,afanasev12} in August and September 2020 (see Table ~\ref{tab:obsopt}) in the dark time (moon phase less than 0.3) and at mean atmosphere seeing less than 2~arcsec. A slit width of ~2~arcsec was used. Technical specifications of the \scorpioii{} spectrograph are described in the user manual.

Spectra were processed using standard IRAF\footnote{http://iraf.noao.edu} mathematical software. The spectral shape was corrected by using observations of spectrophotometric standards from the \citet{massey88} list.

\begin{figure*}
    \centering
    \includegraphics[width=0.9\linewidth]{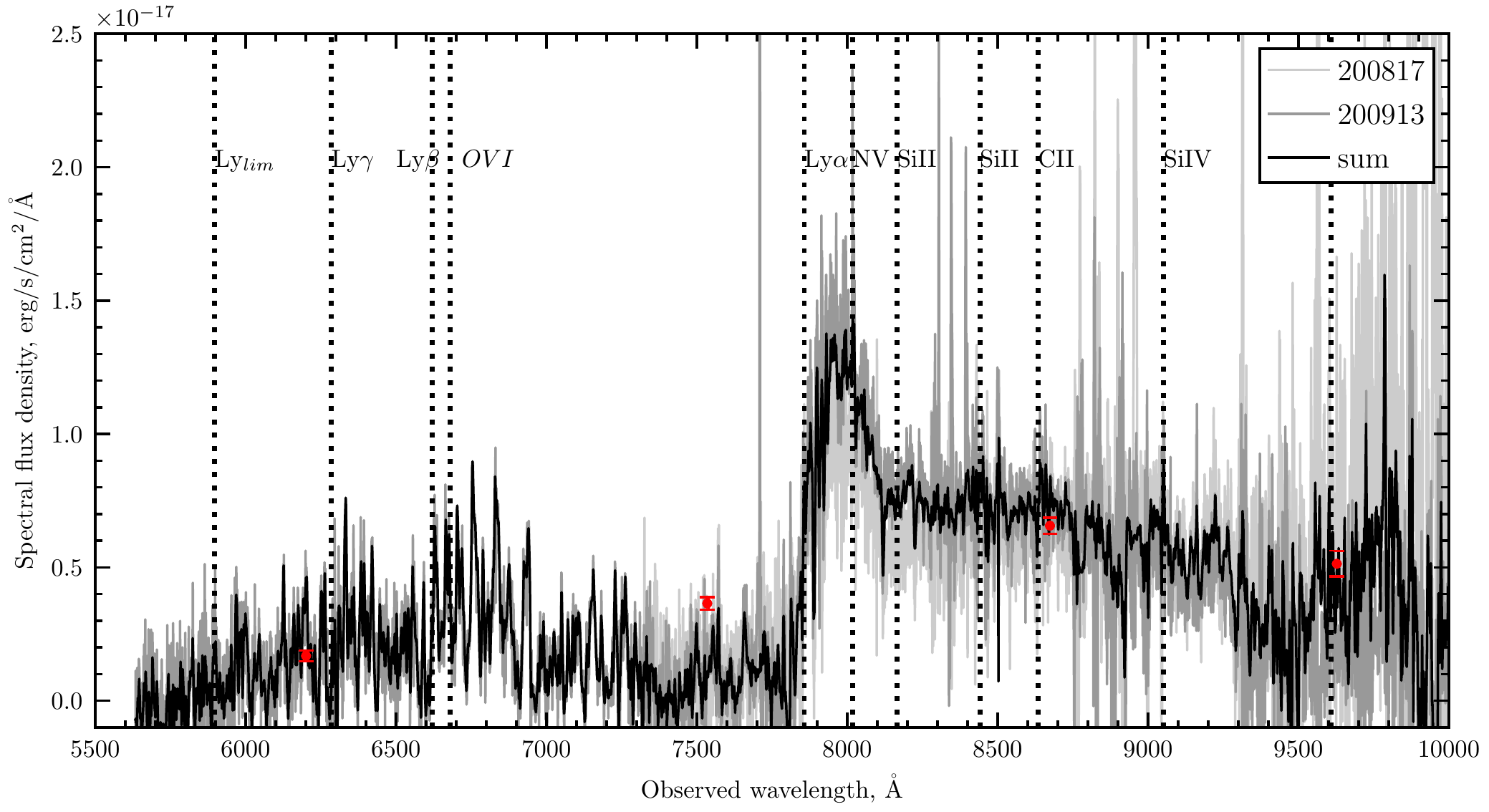}
    
    \caption{Optical spectra of the \theqso quasar obtained with the \BTA telescope. Light gray line shows the spectrum from 2020/08/17. 
Dark gray line shows the spectrum from 2020/09/13. The combined spectrum of the quasar is shown by a black line. The red dots show 
spectral source flux density in the Pan-STARRS $r, i, z, y$ filters. The vertical dashed lines show the expected positions of the peaks of the characteristic emission lines of quasars \citep{vandenberk01} at $\zspec$=5.466.}
    \label{fig:qsospec}
\end{figure*}

In August 2020, the first spectrum of the source was obtained, in which a broad $Ly\alpha$ line with a characteristic break was clearly seen, 
associated with absorption on neutral hydrogen (Fig.~\ref{fig:qsospec}). There are no other emission lines typical for quasars (such as CIV 1549\AA) in the spectrum. 
Sky background lines interfere, and it would take time by a factor of 2 longer to get spectrum of better quality with the VPHG1200@860 grating. 
The $Ly\beta$ line, from which the source redshift could be more accurately determined, does not fall within the operating range of the VPHG1200@860 grating. In September 2020, we repeated spectral observations with the VPHG1026@735 grating, the Lyman-alpha forest can be clearly seen (Fig.~\ref{fig:qsospec}). Unfortunately, the $Ly\beta$ line is very wide, and its peak is in the  absorption. 

\begin{table}[]
\caption{List of BTA observations}
\begin{tabular}{|c|c|c|c|}
\hline
Date & Grating & Exposure time, sec & S/N \\
\hline
 2020/08/17 & VPHG1200@860 & 2x1200  & 3 \\
 
 2020/09/13 & VPHG1026@735 & 5x1200 & 4 \\
 \hline 
\end{tabular}
\label{tab:obsopt}
{\bf Note.} Date --- date of the observation beginning, Grating --- name of the dispersing element, S/N --- average 
signal-to-noise ratio in the spectrum.
\end{table}

According to the results of observations in August and September 2020, the combined spectrum of the quasar \theqso was obtained (shown at Fig.~ref{fig:qsospec}).
This spectrum is similar to the spectra of distant radio-luminosity quasars \citep{romani04, banados18, belladitta20}. Broad emission lines (with exception of $Ly\alpha$) do not dominate the continuum.

There are no narrow emission lines seen in the spectrum of the quasar, which could be used to determine the spectroscopic redshift with higher accuracy. Therefore, the redshift of the source $\zspec=5.466\pm0.003$ was determined by fitting a template of the mean spectrum of the quasar \cite{vandenberk01}. The systematic redshift error is $\Delta \zspec \sim0.01$ and is caused by differences between the source spectrum and the template. The obtained spec-z value is in excellent agreement with the photometric redshift estimate $\zph=5.486$ obtained by \srgz. 

In the quasar spectrum a prominent narrow absorption line feature at a wavelength of 8119 \AA can be seen. Assuming that this is the SiII(1263\AA) line, the other minima of the spectrum (with insufficient detection significance) will agree with the narrow absorption lines ($Ly\alpha$, NV, SiII, SiIV) placed at $z_{abs}=5.427$. Note that in the spectrum of another radio-loud quasar PSO~J352.4034$-$15.3373 at $\zspec=5.84$ a complex of narrow absorption lines at $z_{abs}=5.8213$ \citep{banados18} was also detected. 

\section{Spectral Energy Distribution}

With measured spectoscopic redshift of the quasar \theqso, now we can investigate its spectral energy distribution (SED). We used the following data to construct the SED: Radio --- TGSS \citep{intema17}, NVSS \citep{condon98}, VLASS \citep{gordon20}, IR --- WISE \citep{wright10,dey19,lang16}, Optical --- DESI Legacy Imaging Surveys DR8 (DESI LIS, \cite{dey19}), \panstarrs{} PS1 DR2 stacked \citep{chambers16}.

The resulting SED is shown at Fig.~\ref{fig:sed} (data used for SED construction are given in the Table~\ref{tab:sed}).  Optical and Infrared photometric measurements, as well as luminosities calculated from them, were corrected for Galactic absorption by using the color excess value towards the quasar --- $E(B-V)=0.077$ \citep{schlegel98}. For DESI LIS optical filters and  \wise{} $W1, W2$ filters, the corresponding correction factors mw\_transmission from the DESI LIS DR8 catalog were used. The standard absorption law  \citep{cardelli89} with $R_V=3.1$ was used to correct the Pan-STARRS photometry for Galactic absorption.

\subsection{Radio properties}
\theqso{} is reliably detected in the NRAO VLA Sky Survey (NVSS) at 1.4~GHz (survey observations conducted in the 1990s).
In addition, radio source is registered in the {Very Large Array Sky Survey Epoch 1} (VLASS)\footnote{\url{https://cirada.ca/vlasscatalogueql0}} 
at 2.99~GHz \citep{gordon20} (see Table ~\ref{tab:sed}). One can use the lack of meaningful detection of the \theqso{} in the TGSS survey (survey covered 3.6$\pi$ steradian of the celestial sphere, including the location of the quasar) to put 25~mJy upper limit to the source flux density at $150$~MHz (see \S4.2 \citealt{intema17}). 

Thus, at the moment we have two measurements and one upper limit on the flux density 
in the 150~MHz -- 3 GHz radio range. As can be seen at Fig.~\ref{fig:sed}, according to these data, the radio emission spectrum should have 
maximum between 1 and 10 GHz (in the quasar rest frame). However, this conclusion should be treated with caution because of the probable variability of the object, since the radio observations discussed above were made at different epochs (in particular, NVSS and VLASS measurements are separated by more than 20 years). It is worth noting, that X-ray brightness 
quasar \theqso\ has changed by a factor of 2 at a half-year scale (see"X-ray data"section for more details). 

High source flux density measured in the gigahertz spectral range unequivocally indicates that the quasar \theqso\ is radio-loud. A standard parameter can be used to quantitatively describe radio-loudness:
\begin{equation}
    R\equiv \frac{\fnufiveGHz}{\fnufothsndA},
\end{equation}
where $\fnufiveGHz$ and $\fnufothsndA$ are the flux densities at 5~GHz and at the frequency corresponding 
to the wavelength 4400~\AA\ in the quasar rest frame. To estimate $\fnufiveGHz$ one can assume that the source spectrum around this frequency is described by a power law with slope $\alpha_r=0$ ($S_\nu\propto \nu^{-\alpha_r}$), and normalize 
spectrum by the NVSS measurement at the observed frequency of 1.4~GHz. The value of $\fnufothsndA$ can be calculated  based on the slope $\alpha_{W1,zPS} = 0.2$ of the ultraviolet (in the quasar rest frame) part of the spectrum, determined by the measured flux density values in the $W1$ and $z_{PS1}$ bands (see Table ~\ref{tab:sed}).

Thus, for the quasar \theqso\ we get a value of $R\approx 1200$. Such a high radio luminosity 
is typical for \citep{belladitta20} blazars, i.e., a special category of radio-loud quasars in which the relativistic jet is directed in the observer direction. Only a few objects at $z>5$ are known to have $R>10^3$ \citep{belladitta19}. 
However, if using the method described above to calculate $\fnufiveGHz$ using VLASS data instead of NVSS data, the 
spectral density of $\fnufiveGHz$ will be lower by a factor of 3. As a consequence, the radio loudness estimate will also decrease: 
$R\approx 360$, and $R\sim 100$ values are more typical of ''ordinary'' radio-loud quasars than of blazars 
(see \citealp{belladitta20}). 

It is interesting to compare the SED of the quasar \theqso\ with typical broad-band spectra of radio-loud quasars and blazars. For this purpose at Fig.~\ref{fig:sed} we show the radio-loud quasar average spectrum template from \citep{shang11} and the  blazar average spectrum template (for the radio luminosity range $\log L_{5GHz}=43$--$44$ in which our quasar \theqso falls) from \citep{fossati98}. The radio-loud quasar template was normalized by using stellar magnitude measured for the quasar \theqso{} (in the $z^\prime$ DESI LIS band). 

As can be seen in Fig.~\ref{fig:sed}, the blazar template is poorly suited to describe the SED of the \theqso{} quasar. The template of the radio-loud 
quasar shows significantly better agreement with the observational data. However, it is necessary to take into account the fact that this template is not a universal SED of radio-loud quasars, and the spectral shapefor individual objects may significantly differ  (see \citealp{shang11}).  

The quasar \theqso{} is characterized by the largest measured flux density at 1.4~GHz (26~mJy) among known 
radio-loud quasars (including blazars) at $z\gtrsim5.5$ (see ~Table ~A.1 in \citealp{belladitta20}). In particular, it is nearly twice as bright as the quasar PSO J352.4034$-$15.3373 (14.9~mYan, \citealp{banados18}), which prior to the discovery of the first blazar at $z>6$ (23.7~mYan, \citealp{belladitta20}) was considered as the most powerful radio-luminous quasar at $z>5.5$. It should be noted, however, that at smaller redshifts $4.5<z<5.5$ there are sources with even larger radio-highness values up to $R>10^4$ \citep{belladitta19,kopylov06}.

Concluding the discussion about the radio properties of the quasar \theqso, we note that to clarify whether it is a blazar, it is necessary to conduct a more detailed study of its spectral and spatial characteristics in the radio range. 

\subsection{X-ray/UV luminosity ratio}

The effective spectrum slope ($\aosx$) between 2500~\AA\ and 2~keV \citep{tananbaum79} is often used in studies of quasars:
\begin{equation}
\aosx\equiv -\frac{\log\Big(\Lsx/\Lo\Big)}{\log\Big(\nusx/\nuo\Big)}=-0.3838\,\log\left(\frac{\Lsx}{\Lo}\right),
\label{eq:aosx}
\end{equation}
where $\Lo$, $\Lsx$ --- spectral luminosity density in the quasar rest frame (measured in units of 
erg~s$^{-1}$~Hz$^{-1}$) at 2500~\AA\ and 2~keV, respectively. The parameter $\aosx$ depends 
on relative contribution of different emission mechanisms to the energy release, such as: thermal radiation of accretion disk, re-emission of radiation energy in broad lines,  hot disk corona emission, and radiation from relativistic jets. 

To estimate the $\aosx$ parameter for the \theqso quasar, its monochromatic X-ray luminosity at 2~keV was determined from data of the \erosita telescope, assuming that the X-ray spectrum is described by a power law with slope $\Gamma=1.8$ (see section "X-ray data"). To calculate $\l2500A$, it was assumed that  object optical spectrum can be approximated by radio-loud quasar average spectrum template \citep{shang11} (normalized by the measured apparent magnitude in the $z^\prime$ DESI LIS filter). 
 The derived values of the X-ray and UV luminosities are shown in Table~\ref{tab:sed}. 

\begin{figure*}[htb]
    \centering
    \includegraphics[width=\linewidth]{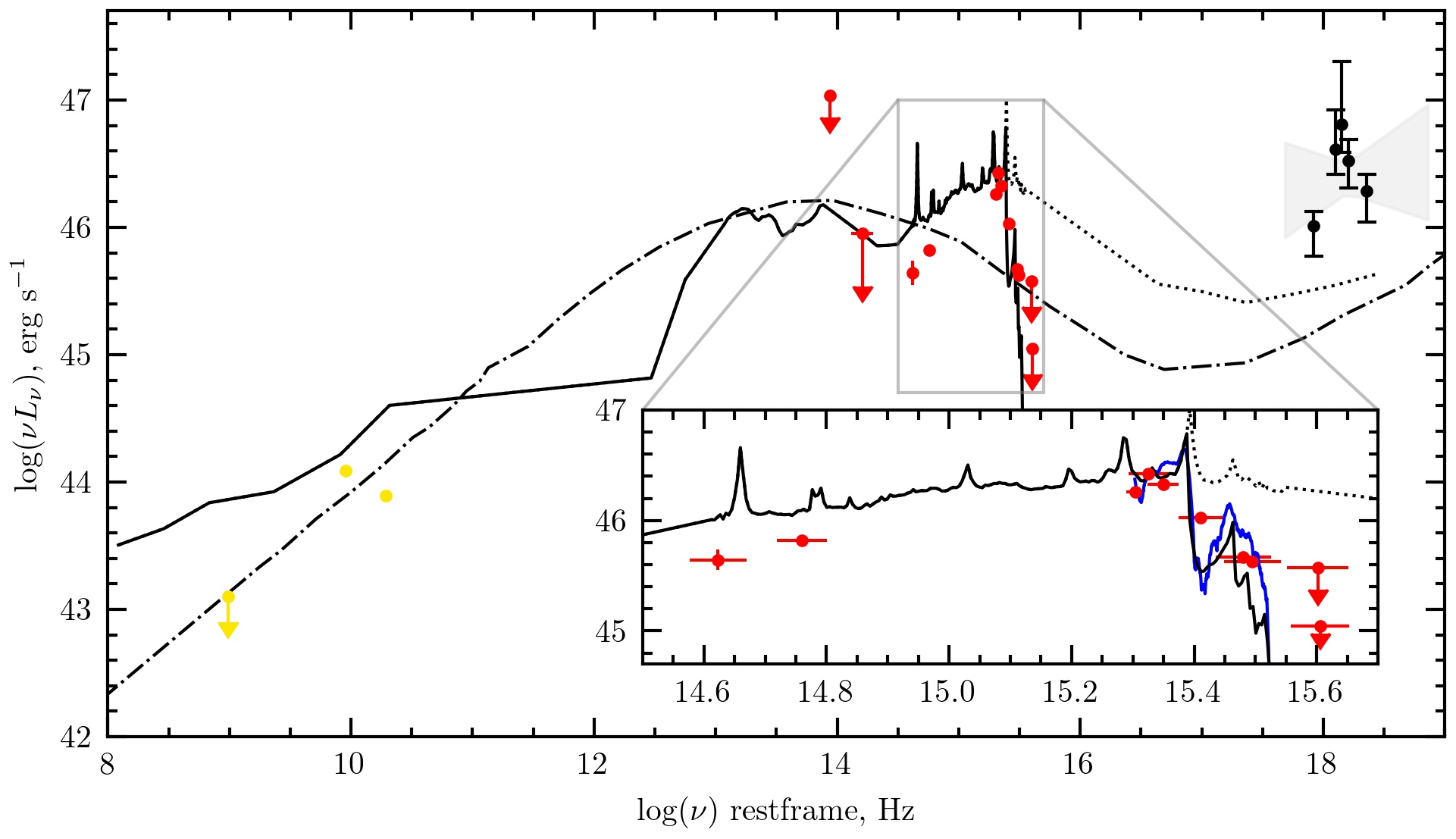}
    \caption{Spectral energy distribution of the \theqso quasar. The yellow dots show measurements in the in the radio spectral range, red dots --- in the near-infrared and visible ranges, black dots --- X-ray data \srge. The gray area --- $1\sigma$ uncertainty of the power law model (with Galactic absorption) of the X-ray spectrum (see text and Table ~\ref{tab:bf}). Dot-dashed line shows the average blazar template \cite{fossati98}. Solid line shows the radio-loud quasar template  \cite{shang11}, corrected at wavelengths $\lambda<1216$\AA \ for neutral hydrogen intergalactic absorption \citep{madau95}; the continuation of the original template \cite{shang11} at wavelengths $\lambda<1216$\AA \ without considering absorption is shown by the dashed line. The blue line in the inset panel shows the smoothed optical spectrum obtained on the BTA telescope. }
    \label{fig:sed}
\end{figure*}

We obtained the following power law slope estimate for \theqso: $\aosx=0.93^{+0.09}_{-0.08}$. This value suggests that the broad-band SED of the quasar has a powerful X-ray excess compared with the vast majority of quasars studied so far. This conclusion is also evidenced by a direct comparison of the \theqso\ SED with average spectra templates for radio-quiet quasars and blazars (see ~Fig.~\ref{fig:sed}).
Radio-quiet quasars are characterized by average slope $\aosx\approx1.37$ \citep{lusso10}. Radio-loud quasars are characterized by slightly higher relative X-ray brightness, but $\aosx\le 1.2$ is extreme for this class of objects as well (see, e.g., \citealp{zhubrandt20}). 

In this sense, the quasar \theqso\ is similar to the quasar \mqso\ = \mqsos\ at $z=6.18$ studied in \cite{medvedev2020}, for which the slope $\aosx=1.11^{+0.25}_{-0.24}$ was obtained. As discussed in \cite{medvedev2020, medvedev21}, 
the powerful X-ray excess in broad-band SED of the quasar \mqso\ may be related to the backward Compton scattering of the cosmic microwave background radiation (whose energy density increases with redshift as $(1+z)^4$) in relativistic jets. It is possible that we observe a similar phenomenon in a slightly closer quasar \theqso. 

\begin{figure*}[htb]
    \centering
    \includegraphics[width=1.0\linewidth]{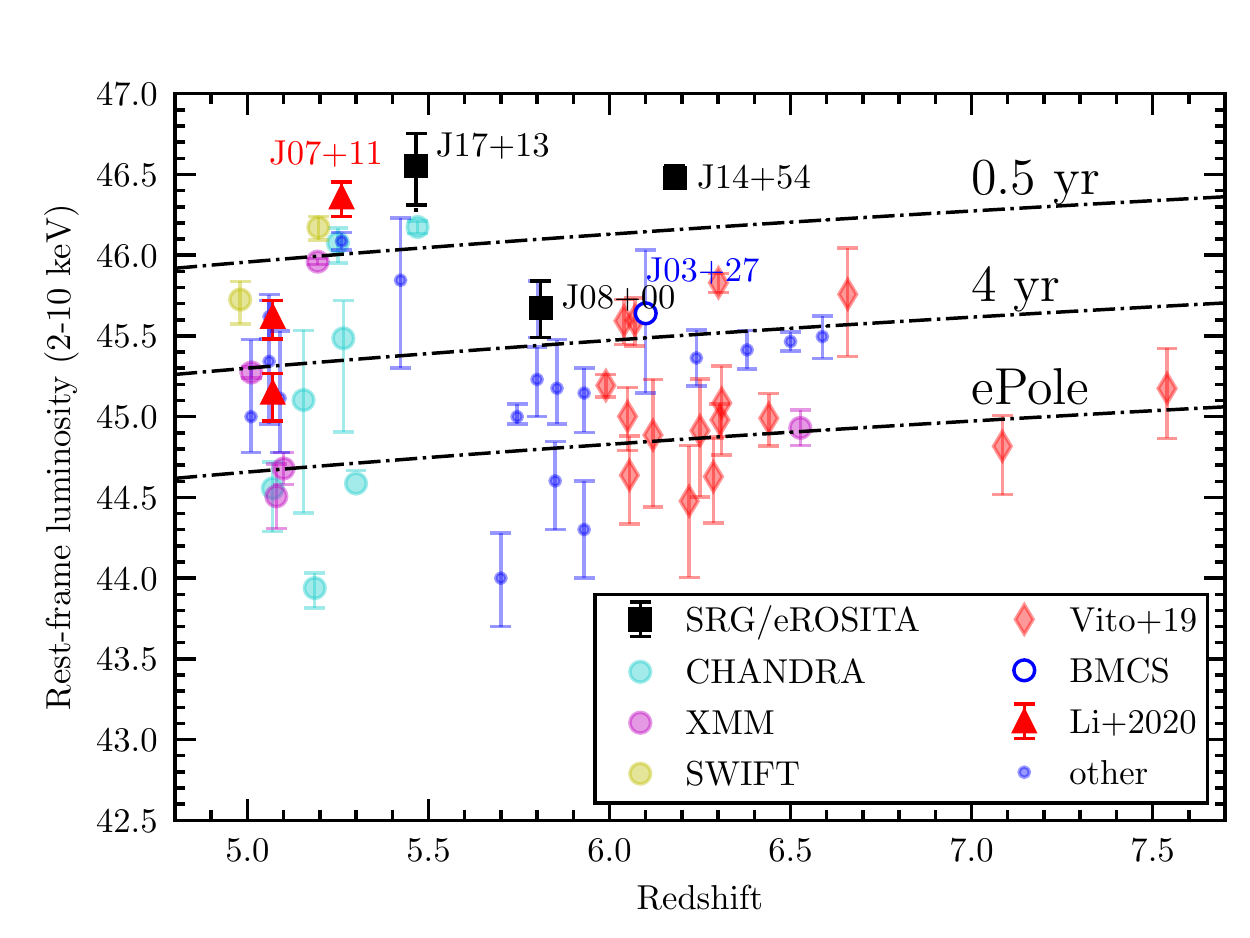}
    \caption{X-ray luminosity distribution of known quasars at $z>5$, which have been registred in X-rays by the \chandra, \xmm, Neil Gehrels Swift Observatory (including the three radio-quiet quasars from the \cite{liwang20} article.  The black squares show prominent quasars detected in X-rays by the \srg\ observatory: \theqso (this paper), \mqso\ \citep{medvedev2020} and SDSS~J083643.85+005453.3 \citep{wolf21} (the first two objects --- during the first scan of All-Sky Survey, the third object --- from the deep survey of the eFEDS field carried out during the verification phase of the observatory).
     For the quasars \theqso\ and SDSS~J074749.18+115352.46 (J07+11) a significant X-ray variability is detected, thus only luminosity values in the brighter source state are shown. The most distant blazar known from literature (BMCS, \cite{belladitta20}) is labeled by "J03+27" mark. Red triangles with the apex pointing downward show sources from the \citep{vito19} catalog of X-ray quasars at $z>6$. Dashed lines denote the characteristic sensitivity thresholds \srge for: 1 scan survey (0.5 yr), 8 scans survey (4 yr), and for regions near the ecliptic poles (ePole) \citep{predehl20, sunyaev20}.}
    \label{fig:lxray}
\end{figure*}

\subsection{Bolometric luminosity}

On the basis of the broad-band SED presented at Fig.~\ref{fig:sed} we can estimate the bolometric luminosity of the quasar \theqso in a similar was as it was done in \citep{medvedev2020} for another distant radio-loud quasar \mqso\ = \mqsos. 

Using the radio-loud quasar template from \citep{shang11}, we can estimate the luminosity at energies below 2~keV. It is $L_{<2~{\rm keV}}\approx 1.2 \times 10^{47}$~erg/s. The luminosity in the 2--100~keV range can be estimated  by extrapolating the X-ray spectrum with slope $\Gamma = 1.8$, measured (at energies 2--32~keV in the 
quasar rest frame) with the \erosita{} telescope. Thus, we obtain $L_{2-100~{\rm keV}} = (7 \pm 3 ) \times 10^{46}$~erg/s. 
Assuming that the contribution from higher energies (above 100~keV) is negligible, we find that the bolometric luminosity of the quasar \theqso\ is $L_{\rm bol}=$ (1.5--2)$\times 10^{47}$~erg/s. 

Assuming that the bolometric luminosity does not exceed the Eddington critical luminosity, we can obtain a lower limit on the mass of a supermassive black hole in the \theqso quasar as $\Mbh> 10^9 \Msun$. 

Thus, the bolometric luminosity and black hole mass of the \theqso\ quasar appear to be comparable to the corresponding characteristics of other brightest radio-loud quasars at $z>5$ (see,~e.g.,~\cite{medvedev2020}).

\section{\theqso \ ---the brightest X-ray quasar at $z>5$}

Discovered by \erosita\ telescope of \srg\ observatory, quasar \theqso\ has enormous X-ray luminosity and at the moment is the brightest in X-rays among all known quasars at $z>5$ at Eastern Galactic Hemisphere (processing data \srge\ on which the Russian scientists are responsible). Is such a high X-ray luminosity unique at all for quasars at $z>5$?

To answer this question, we've cataloged all the X-ray registered quasars at the moment 
at $z>5$. Details of how this information was collected are described in the Appendix at the end of the paper. The resulting catalog (a total of 52 objects) is presented in Table ~\ref{tab:xzcat}, references to the X-ray data and redshifts are given in Tables ~\ref{tab:xref} and \ref{tab:oref}, respectively. Based on the information collected. 
it is possible to compare the X-ray luminosities of the \theqso\ quasar and previously known quasars. The results of this comparison are presented in Fig.~\ref{fig:lxray}. It can be seen that the radio-loud quasar \theqso\ is the brightest in X-ray luminosity among all known quasars at $z>5$ (including blazars).  

Note that in the second scan of \srge\ All-Sky Survey (six months after the first observation) X-ray luminosity of the quasar \theqso\ has decreased by about factor of 2 (see section "X-ray data"). It is worth noting, that the previously known brightest X-ray radio-loud quasar SDSS~074749.18+115352.46 exhibits X-ray variability of comparable 
amplitude on a scale of few hours \citep{liwang20}.

Fig.~\ref{fig:lxray} shows that X-ray observatories so far has been focused mainly on quasars at $z\gtrsim6$. The unique data from the four-year \srge survey will reveal all objects in the poorly understood part of quasar luminosity function at extremely large X-ray luminosities ($L_{X}>5\times10^{45}$~erg/s). The combined use of the data from X-ray and Optical sky surveys will help improve the purity of distant quasars candidate samples selection and allow 
to fill the apparent gap in the redshift distribution of luminous quasars (~Fig.~\ref{fig:lxray}).

\section{Conclusion}

The X-ray quasar \theqso at $z\approx 5.47$ discovered by the \srg{} X-ray observatory and the \BTA{} 6-m telescope is the most X-ray luminous quasar among known objects in the early Universe ($z>5$) and also one of the most powerful quasars in radio.

The large radio-loudness ($R\sim 10^3$) of the quasar indicates that it can be a blazar. To test this hypothesis, it is necessary to carry out interferometric radio observations of the object at several wavelengths. Note that only a few blazars at $z>5$ \citep{belladitta20} are currently known, and all of them have lower X-ray luminosities than the quasar \theqso.

Spectroscopic measurements in the near-infrared range ($\lambda \sim1.6$~microns) could also significantly complement the physical picture. It is expected that a broad MgII emission line should appear in the near-IR, which could be used to measure the mass of the black hole.

The quasar \theqso\ shows significant X-ray variability in the first two scans of the \erosita\ all-sky survey. We will continue to monitor its variability in the following scans. 

The X-ray detection of the radio-loud quasars \mqso\ = \mqsos\ at $z=6.18$ \citep{medvedev2020} 
and \theqso\ at $z=5.47$ with record-breaking X-ray luminosities of $\sim 3\times 10^{46}$~erg/s by the \srg{} observatory opens a new page in the study of supermassive black hole growth in the early Universe. 
We hope that the sample of such interesting objects will be substantially expanded during the ongoing all-sky X-ray survey. Spectroscopic verification of new X-ray sources plays a key role. Searches for unique quasars among the sources discovered by the \srg \ observatory are continuing with the help of the BTA 6-m  telescope, AZT-33IK 1.6-m telescope \citep{burenin16} and the Russian-Turkish RTT-150 1.5-m telescope \citep{bikmaev20}.

\paragraph{}  
Observations on the telescopes of the SAO RAS are carried out 
with the support of the Ministry of Science and Higher Education of the Russian Federation (Ministry of Education and Science of Russia). 

This study is based on observations of the telescope \erosita\ aboard the \srg observatory. The observatory was built by Roscosmos on behalf of the Russian Academy of Sciences, represented by the Institute of Space Research (IKI) within the framework of the Russian Federal Science Program with participation of the German Center for Aeronautics and Astronautics (DLR). The X-ray telescope \srge\ was made by a consortium of German institutes led by the Max Planck Society Institute for Extraterrestrial Astrophysics (MPE) with the support of DLR. The spacecraft \srge\ is designed, manufactured, launched, and operated by the Lavochkin Research Center and its subcontractors. Scientific data are received by a complex of long-range space communications antennas in Bear Lakes, Ussuriysk and Baikonur and are funded by Roscosmos. The eRosita telescope data used in this work are processed using the eSASS software developed by the German consortium \erosita\ and the software developed by the Russian telescope consortium \srge. The \srgz{} system was created in the Science Working Group of RU \erosita{} consortium on X-ray  source  detection,  identification  and  \srge source catalog. 

The functions from the astropy\citep{robitaille13}, pysynphot\citep{lim15}, and specutils\footnote{\url{https://pysynphot.readthedocs.io}} libraries were used to calculate Galactic\footnote{\url{https://extinction.readthedocs.io/en/latest/}} and extragalactic\footnote{\url{https://specutils.readthedocs.io/} 
absorption, distances and other astrophysical quantities.} The transmission curves and other characteristics of the photometric filters are taken from the Spanish Virtual Observatory Filter Profile Service\footnote{\url{http://svo2.cab.inta-csic.es/theory/fps/}} \cite{rodrigo12,rodrigo20}. The catalogs of the VizieR database \cite{ochsenbein00} were used in this work.

This study was supported by the Russian Science Foundation (grant 19-12-00396).

\section{\it Appendix: Catalogue of X-ray quasars at $z>5$}

Commonly known and widely used samples of distant ($z\gtrsim 5.5$) X-ray quasars were compiled several years ago (see, ~e.g., \citealp{nanni17, vito19}). The search for new quasars is actively ongoing in all bands of the electromagnetic spectrum, so published catalogs are rapidly becoming outdated. In light of the successful work of the 
\srg\ observatory, we have prepared our updated and complete (at December 2020) catalog of all spectroscopically confirmed quasars at $z>5$, registered in the X-ray range (Table ~\ref{tab:xzcat}).

The catalog is based on a joint sample of spectroscopically confirmed quasars at $z>5$ from "The Million Quasars catalog, v7.0a"\\citep{flesch19} and from "VHzQ"\ $z>5$ quasars catalog  \citep{ross20}. The later sample (contains 542 "optical"\ quasars at $z\gtrsim5$) was correlated with the list of X-ray sources optical counterparts taken from the list of articles given in Table$\ref{tab:xref}$. Information on the X-ray fluxes and luminosities of the X-ray sources is given in the Table ~$\ref{tab:xzcat}$, together with information about coordinates of the optical counterparts and their spectral redshifts (from \cite{flesch19}, \cite{ross20}). The observed X-ray flux of 0.5--2~keV is given for most objects. 
As in the \cite{salvestrini19} paper there was no information about X-ray fluxes of sources, we also does not provide this information in our Table~\ref{tab:xzcat}. In the REF($FX$) column for each source we give reference to the paper from which the X-ray fluxes are taken, and than mention all the articles with X-ray observations of the source. Luminosities are given in the range 2--10~keV in the quasar reference frame. 
The values of the X-ray flux and the luminosity for sources from \citep{vito19} are taken mainly from Table~7 of the \cite{vito19}.

For the remaining $z>5$ optical quasars, for which no X-ray companion was found in the literature, we searched X-ray catalogs for the nearest object in 6~arcsec aperture radius. We used the following X-ray catalogs: Chandra source catalog Release 2.0 \citep{evans10,evans20}, \xmm 4XMM-DR10 Catalog \citep{webb20}, 2SXPS Swift X-ray telescope Point source catalog \citep{evansp20}. The X-ray luminosities $\Lnx$ were calculated in the 2--10~keV energy range (in the quasar rest frame). This assumed a power law spectrum with $\Gamma=1.8$ and used measurements from the \chandra\ and \xmm\ observatories in the 0.5--2~eV range and data from the \swift\ observatory in the 0.3--10~eV range. 

As a result, a catalog of 52 X-ray quasars at $z>5$ was obtained, including the quasar \theqso discussed in this paper. 
Table ~\ref{tab:oref} contains references to the articles with spectroscopic redshifts for these objects.

\vfill
\eject

\onecolumn
\landscape
\input{4sed_table}

\clearpage
\newpage
\input{5cattab}

\end{document}

%% file: 4sed_table.tex
\begin{table*}
  \caption{Multi-wavelength properties of the quasar \theqso {\bf 
  }
    \label{tab:sed}
  }
  \begin{tabular}{|lllrrr|}
    \hline
    %
    \multicolumn{1}{|c|}{Telescope/Survey} &    
    \multicolumn{1}{|c|}{Passband} &
    \multicolumn{1}{c|}{Flux or magnitude} &
    \multicolumn{1}{c|}{Reference} & 
    \multicolumn{1}{c|}{$\nu$ in rest frame (Hz)} & 
    \multicolumn{1}{c|}{$\nu L_{\nu}$, erg~s$^{-1}$} \\
    \hline
    \multicolumn{6}{|c|}{Radio range:}\\
    TGSS & 150~MHz & $<$25 mJy           & 1 & $9.70\times 10^{8}$ & $<1.3\times10^{43}$\\
    NVSS & 1.4~GHz & $26.0\pm 0.9$ mJy & 2 & $9.05\times 10^{9}$ & $(1.22\pm0.04) \times 10^{44}$ \\
    VLASS & 2.99~GHz & $7.8\pm0.3$ mJy & 3 & $1.93\times 10^{10}$ & $(7.86\pm0.32) \times 10^{43}$  \\
    \hline
    \multicolumn{6}{|c|}{Infrared range:}\\
    WISE/PS1 & $W2_{\rm Vega, forced}$  & $17.05\pm0.13$ & 4 & $4.198\times10^{14}$ & $(5.50\pm 0.67)\times 10^{45}$ \\
    & $W1_{\rm Vega, forced}$ & $17.82\pm0.07$ & & $5.755\times10^{14}$ & $(6.71\pm 0.43)\times 10^{45}$ \\
    WISE/DESI LIS & W2$_{\rm Vega, forced}$ & $17.30\pm0.23$ & 5 & $4.198\times10^{14}$ & $(4.39\pm 0.94)\times 10^{45}$\\ 
    & W1$_{\rm Vega, forced}$ & $17.84\pm0.09$ & &$5.755\times10^{14}$ & $(6.60\pm 0.56)\times 10^{45}$\\ 
    \hline
    \multicolumn{6}{|c|}{Optical range:}\\
    PS1 & $y$ & $20.80\pm0.10$ & 6 & $2.014\times10^{15}$ & $(1.82\pm 0.17)\times 10^{46}$ \\
    & $z$ & $20.74\pm0.05$ & & $2.235\times10^{15}$ & $(2.13\pm 0.10)\times 10^{46}$ \\
    & $i$ & $21.64\pm0.07$ & & $2.573\times10^{15}$ & $(1.07\pm 0.07)\times 10^{46}$ \\
    & $r$ & $22.86\pm0.13$ & & $3.126\times10^{15}$ & $(4.24\pm 0.50)\times 10^{45}$ \\
    & $g$ & $<23.25$ & & $3.998\times10^{15}$ & $<3.8\times 10^{45}$\\
    DESI LIS & $z^\prime$, & $20.44\pm0.02$ & 4 & $2.114\times10^{15}$ & $(2.66\pm0.05)\times10^{46}$\\
    & $r^\prime$, & $22.71\pm0.06$ & & $3.021\times10^{15}$ & $(4.69\pm0.25)\times10^{45}$\\
    & $g^\prime$,  & $<24.58$ & & $4.031\times10^{15}$ &$<1.1\times10^{45}$\\
    \hline
    \multicolumn{6}{|c|}{X-ray range:}\\
    SRG/eROSITA & 0.5--2~keV ($\Gamma=1.8$) & $1.0^{+0.5}_{-0.4}\times 10^{-13}$~erg~cm$^{-2}$~s$^{-1}$ & 4 & & \\
    \hline
    \multicolumn{6}{|l|}{Derived values:} \\
    \multicolumn{2}{|l}{$\aosx$ (2500~\AA--2~keV)} & $0.93^{+0.09}_{-0.08} $& & & \\
    \multicolumn{2}{|l}{$\alpha_{W1,zPS}$} & $0.20\pm0.02$ & & & \\
    \multicolumn{2}{|l}{$\nu L_{\nu}(5GHz)$} & $(6.7\pm0.2) \times 10^{43} $~erg~s$^{-1}$  & & & \\
    \multicolumn{2}{|l}{$\nu L_{\nu}$(4400~\AA)} & $(7.68\pm0.03) \times 10^{45} $~erg~s$^{-1}$  & & & \\
    \multicolumn{2}{|l}{$\nu L_\nu$ (2500~\AA)} & $ (1.20 \pm 0.02) \times 10^{46}$~erg~s$^{-1}$ & & & \\

\multicolumn{2}{|l}{$\nu L_\nu$ (2~keV)} & $1.9^{+1.1}_{-0.8} \times 10^{46}$~erg~s$^{-1}$ & & & \\
    \multicolumn{2}{|l}{$L$ (bolometric)} & (1.5--2) $\times 10^{47}$~erg~s$^{-1}$& & & \\
    \hline
\end{tabular}


\textbf{Note.} Luminosities are given in the quasar rest frame and corrected for Galactic extinction.\\
\textbf{References:} (1) \cite{intema17}, (2)  \cite{condon98}, (3) \cite{gordon20}  (4) this article, (5) \cite{dey19} DESI LIS DR8, (6) \cite{chambers16} PS1 DR2 stacked

\end{table*}

%% file: 5cattab.tex
\begin{longtable}{lD{.}{.}{3.4}D{.}{.}{3.4}cccllll}
    \label{tab:xzcat}\\
    \caption{Catalog of X-ray quasars at $z>5$}\\
    \hline
    \multicolumn{1}{|c|}{Source} &    
    \multicolumn{1}{|c|}{RA} &  
    \multicolumn{1}{|c|}{DEC} &  
    \multicolumn{1}{|c|}{$z$} &
    \multicolumn{1}{c|}{REF($z$)} &
    \multicolumn{1}{c|}{$z_{RC20}$} & 
    \multicolumn{1}{c|}{$F_X$} & 
    \multicolumn{1}{c|}{REF($F_X$)} & 
    \multicolumn{1}{c|}{$log(L_{X,2-10})$} &
    \multicolumn{1}{c|}{REF($L_X$)}\\

    \multicolumn{1}{|c|}{} &    
    \multicolumn{1}{|c|}{deg} &    
    \multicolumn{1}{|c|}{deg} &
    \multicolumn{1}{|c|}{} &
    \multicolumn{1}{c|}{} &
    \multicolumn{1}{c|}{} & 
    \multicolumn{1}{c|}{$\times10^{-14}$~erg/s/cm$^2$} & 
    \multicolumn{1}{c|}{} & 
    \multicolumn{1}{c}{erg/s} &
    \multicolumn{1}{|c|}{} \\
    \hline
\endhead

\hline
\hline
\endlastfoot
\input{5Re_qso5catalogXray.tex}
\end{longtable}

\textbf{Note.} Source --- name of the quasar, RA, DEC --- Right Ascention and Declination (J2000) of the optical companion, $z$ --- spectroscopic redshift, REF($z$) --- bibliographic reference to the $z$ measurement (see Table~\ref{tab:oref} Table~\ref{tab:oref} below), $z_{RC20}$-- spectroscopic redshift from the \cite{ross20} catalog, $F_X$-- X-ray flux (symbol "*" denotes that the flux value is taken from a subsidiary reference catalog). REF($F_X$) --- bibliographic reference to the $F_X$ value (the range of the measured $F_X$ is due to the corresponding X-ray survey, see ''The X-ray Survey'', ''\theqso --- the brightest X-ray quasar at $z>5$'' sections and Table~\ref{tab:xref}), $\log(L_{X,2-10})$ --- X-ray luminosity in the 2--10~keV range in the quasar rest frame. The luminosity calculation was performed only for objects with REF($L_X$) $=1,2,3$. Spectroscopic values $z$ were used in the luminosity calculation, $k$-correction was calculated for the power X-ray spectrum with photon index $\Gamma=1.8$ without taking into account Galactic/extragalactic absorption. For other sources (REF($L_X$)$>3$) the values $F_X$ and $\log(L_{X,2-10})$ are given from original articles (see relevant references). 

\newpage

\begin{table}
   \caption{Bibliographic references for X-ray fluxes and luminosities from Table~5}
    \label{tab:xref}
    \centering
    \begin{tabular}{|c|c|c|}
    \hline
    \multicolumn{1}{|c|}{REF($FX$), REF($LX$)} &
    \multicolumn{1}{c|}{Reference} &
     \multicolumn{1}{c|}{X-ray range, keV} \\
    \hline
        1 & \cite{evans20} & 0.5-2 \\
        2 & \cite{webb20} & 0.5-2 \\
        3 & \cite{evansp20} & 0.3-10\\
        4 & \cite{civano16} & 0.5-2\\
        5 & \cite{vito19} & 0.5-2\\
        6  & \cite{liwang20} & 0.5-2 \\
        7  & This article & 0.5-2 \\
        8 & \cite{medvedev21} & 0.2-10 \\
        9 &  \cite{wolf21} &  0.5-2  \\
       10 &  \cite{nanni17} & 0.5-2 \\
       11 &  \cite{wang21} & 0.5-2 \\     
       12 & \cite{connor19} & 0.5-2 \\
       13 & \cite{pons20} & 0.5-2 \\
       14 &  \cite{salvestrini19} & --- \\
       15 &  \cite{belladitta20} & 0.3-10 \\
    \hline
    \end{tabular}
   
\end{table}

\begin{table}
   \caption{Bibliographic references for spectroscopic redshifts from Table~5}
    \label{tab:oref}
    \centering
    \begin{tabular}{|c|c|}
    \hline
    \multicolumn{1}{|c|}{REF($z$)} & 
    \multicolumn{1}{c|}{Reference} \\
    \hline
127&\cite{barger02}\\ 
144&\cite{becker01} \\ 
588&\cite{fan03}\\ 
589&\cite{fan04}\\ 
590&\cite{fan06} \\ 
695&\cite{goodrich01}\\ 
1139&\cite{mahabal05}\\ 
1144&\cite{maiolino04}\\ 
1221&\cite{masters12} \\ 
1244&\cite{mcgreer09}\\ 
1245&\cite{mcgreer13} \\ 
1318&\cite{mortlock11}\\ 
1457&\cite{pettini03}\\ 
1557&\cite{romani04}\\ 
2049&\cite{willott07}\\ 
2051&\cite{willott10}\\ 
BMCS&\cite{belladitta20}\\ 
DR14Q&\cite{paris18}\\ 
DR16Q&\cite{lyke20}\\ 
HIZ7.5&\cite{banados18b}\\ 
KHOR1&\cite{khorunzhev17}\\ 
KHOR21&This article\\ 
OVRLAP&\cite{jiang15}\\ 
PS1&\cite{banados16}\\ 
PS1MAZ&\cite{mazzucchelli17}\\ 
PSO&\cite{venemans15}\\ 
ULTRA&\cite{wu15}\\ 
VAHIZ&\cite{carnall15}\\ 
VDES&\cite{reed17}\\ 
WISEHI&\cite{wang16}\\ 
\hline
    \end{tabular}
   
\end{table}

%% file: 5Re_qso5catalogXray.tex
SDSS J00026+2550&0.6642&25.8430&5.800&0589&5.820&0.39$_{-0.16}^{+0.24}$&10&45.23$_{-0.23}^{+0.20}$&10\\
SDSS J000552.33-000655.6&1.4681&-0.1155&5.855&DR16Q&5.850&0.25$_{-0.19}^{+0.27}$&10,1&45.18$_{-0.22}^{+0.30}$&10\\
SDSS J002526.84-014532.5&6.3618&-1.7590&5.060&DR16Q&5.070&1.41$_{-0.38}^{+0.37}$&6&45.62$_{-0.14}^{+0.10}$&6\\
CFHQS J0050+3445&12.5278&34.7563&6.250&2051&6.253&0.145$_{-0.089}^{+0.157}$&5&44.91$_{-0.41}^{+0.32}$&5\\
SDSS J010013.02+280225.8&15.0542&28.0405&6.301&ULTRA&6.326&0.771$_{-0.102}^{+0.110}$&5,10,2&45.83$_{-0.06}^{+0.06}$&5\\
HRQC J011544.78+001514.9&18.9366&0.2541&5.100&1245&5.100&0.154$\pm0.039$&2&44.68$\pm0.10$&2\\
SDSS J013127.34-032100.1&22.8639&-3.3500&5.196&DR16Q&5.180&13.18$_{-2.17}^{+2.17}$&3&46.17$_{-0.08}^{+0.07}$&3\\
ATLAS J025.6821-33.4627&25.6821&-33.4627&6.310&VAHIZ&6.338&0.198$_{-0.103}^{+0.143}$&5,10,3&45.08$_{-0.32}^{+0.23}$&5\\
SDSS J022112.62-034252.2&35.3026&-3.7145&5.011&DR16Q&5.020&0.628$\pm0.052$&2&45.27$\pm0.03$&2\\
VDESJ0224-4711&36.1106&-47.1915&6.500&VDES&6.500&0.508$\pm0.065$&13,11,2&45.47$\pm0.06$&13\\
PSO J036.5078+03.0498&36.5078&3.0498&6.527&PSO&6.541&0.161$\pm0.046$&2,11&44.93$\pm0.11$&2\\
SDSS J023137.64-072854.4&37.9069&-7.4818&5.423&DR16Q&5.370&1.925*$_{-0.649}^{+0.633}$&14,2,1*&45.85$_{-0.15}^{+0.15}$&14\\
PSO J030947.49+271757.31&47.4479&27.2993&6.100&BMCS&&3.4$_{-1.9}^{+5.2}$&15&45.64$_{-0.50}^{+0.39}$&15\\
SDSS J074154.72+252029.6&115.4780&25.3416&5.194&1244&5.194&2.818$\pm0.126$&2,1&45.96$\pm0.02$&2\\
SDSS J074749.18+115352.4&116.9549&11.8979&5.260&WISEHI&5.260&3.49$_{-0.45}^{+0.44}$&6&46.09$\pm0.05$&6\\
SDSS J075618.13+410408.6&119.0756&41.0691&5.060&DR16Q&5.060&0.64*$_{-0.29}^{+0.28}$&14,1*&45.34$_{-0.23}^{+0.20}$&14\\
SDSS J08367+0054&129.1831&0.9147&5.803&1457&5.810&9.9$_{-3.2}^{+3.7}$&9,10,14,1&45.67$_{-0.18}^{+0.17}$&9\\
SDSS J084035.09+562419.9&130.1463&56.4056&5.850&0590&5.844&0.09$_{-0.05}^{+0.07}$&10,1&44.60$_{-0.30}^{+0.24}$&10\\
SDSS J084229.43+121850.5&130.6226&12.3140&6.055&OVRLAP&6.076&0.075$_{-0.038}^{+0.056}$&5&44.64$_{-0.30}^{+0.24}$&5\\
Q J0906+6930&136.6283&69.5086&5.470&1557&5.470&4.127$_{-0.355}^{+0.355}$&1,3&46.17$_{-0.04}^{+0.04}$&1\\
COSM J095908.1+022707&149.7838&2.4521&5.070&1221&5.070&0.117$_{-0.054}^{+0.053}$&1&44.56$_{-0.27}^{+0.16}$&1\\
COSM J100051.6+023457&150.2150&2.5827&5.300&1221&5.300&0.114$\pm0.024$&4&44.59$\pm0.08$&4\\
SDSS J102623.62+254259.4&156.5985&25.7165&5.250&DR16Q&5.250&3.599$_{-0.903}^{+0.854}$&1,3&46.08$_{-0.13}^{+0.09}$&1\\
SDSS J10304+0524&157.6131&5.4153&6.305&1144&6.308&0.176$_{-0.038}^{+0.044}$&5,10,14,1,2&44.98$_{-0.10}^{+0.10}$&5\\
PSO J159.2257-02.5438&159.2258&-2.5439&6.380&PS1&6.381&0.411$\pm0.058$&13,2&45.41$\pm0.12$&13\\
SDSS J10445-0125&161.1381&-1.4172&5.745&0695&5.785&0.31$_{-0.04}^{+0.05}$&10,2&45.00$_{-0.05}^{+0.08}$&10\\
SDSS J10487+4637&162.1877&46.6218&6.287&DR14Q&6.228&0.077$_{-0.037}^{+0.056}$&5,10&44.63$_{-0.29}^{+0.24}$&5\\
SDSS J105036.46+580424.6&162.6520&58.0735&5.155&DR16Q&5.155&0.398$_{-0.318}^{+0.677}$&1&45.10$_{-0.70}^{+0.43}$&1\\
SDSS J105322.98+580412.1&163.3458&58.0700&5.265&DR16Q&5.265&0.917$_{-0.676}^{+0.655}$&1&45.49$_{-0.58}^{+0.23}$&1\\
ULAS J112001.48+064124.3&170.0062&6.6901&7.085&1318&7.084&0.068$_{-0.028}^{+0.048}$&5,11,10,14,1,2&44.82$_{-0.30}^{+0.19}$&5\\
SDSS J114657.79+403708.6&176.7408&40.6191&4.980&DR16Q&5.009&5.167$_{-1.526}^{+1.526}$&3&45.73$_{-0.15}^{+0.11}$&3\\
RD J1148+5253&177.0675&52.8942&5.700&1139&5.700&0.02$_{-0.01}^{+0.02}$&10&44.00$_{-0.30}^{+0.28}$&10\\
SDSS J114816.64+525150.3&177.0694&52.8640&6.440&DR16Q&6.419&0.196$_{-0.064}^{+0.083}$&5,10,14&44.99$_{-0.17}^{+0.15}$&5\\
SDSS J120441.73-002149.6&181.1739&-0.3638&5.090&DR16Q&5.090&0.40*$_{-0.23}^{+0.23}$&14,1*&45.11$_{-0.27}^{+0.21}$&14\\
B01.174&189.1998&62.1615&5.186&0127&5.186&0.027$_{-0.007}^{+0.006}$&1&43.94$_{-0.12}^{+0.09}$&1\\
3XMM J125329.4+305539&193.3721&30.9277&5.080&KHOR1&5.080&0.104$\pm0.061$&2&44.51$\pm0.20$&2\\
SDSSp J130608.26+035626.3&196.5344&3.9406&5.990&0144&6.034&0.322$_{-0.049}^{+0.544}$&5,10,14,1&45.19$_{-0.07}^{+0.07}$&5\\
SDSS J13358+3533&203.9617&35.5544&5.930&0590&5.901&0.04$_{-0.02}^{+0.04}$&10&44.30$_{-0.30}^{+0.30}$&10\\
ULAS J134208.10+092838.6&205.5337&9.4774&7.540&HIZ7.5&7.540&0.173$_{-0.088}^{+0.133}$&5,11&45.17$_{-0.31}^{+0.25}$&5\\
SDSS J14111+1217&212.7972&12.2936&5.930&0589&5.904&0.35$_{-0.20}^{+0.23}$&10,14,1&45.15$_{-0.24}^{+0.15}$&10\\
CFHQS J1429+5447&217.4674&54.7882&6.210&2051&6.183&10.8$_{-1.0}^{+0.9}$&8&46.48$_{-0.06}^{+0.08}$&8\\
CFHQS J15096-1749&227.4242&-17.8242&6.120&2049&6.122&0.142$_{-0.091}^{+0.172}$&5&44.89$_{-0.45}^{+0.34}$&5\\
SDSS J16029+4228&240.7256&42.4731&6.070&0589&6.090&0.689$_{-0.210}^{+0.262}$&5,10,14,1&45.60$_{-0.16}^{+0.14}$&5\\
SDSS J16235+3112&245.8831&31.2003&6.220&0589&6.260&0.089$_{-0.059}^{+0.107}$&5,10,1&44.48$_{-0.47}^{+0.34}$&5\\
SDSS J16305+4012&247.6414&40.2028&6.050&0588&6.065&0.204$_{-0.787}^{+0.105}$&5,10,14,1,2&45.00$_{-0.21}^{+0.18}$&5\\
CFHQS J16413+3755&250.3405&37.9223&6.040&2049&6.047&0.636$_{-0.181}^{+0.226}$&5&45.59$_{-0.15}^{+0.13}$&5\\
SRGE J170245.3+130104&255.6888&13.0173&5.466&KHOR21&&10.3$_{-4.0}^{+5.1}$&7&46.56$_{-0.23}^{+0.19}$&7\\
PSO J308.0416-21.2339&308.0416&-21.2340&6.240&PS1&6.234&0.43*$\pm0.09$&12,2*&45.36$_{-0.17}^{+0.17}$&12\\
PSO J323.1382+12.2986&323.1383&12.2987&6.588&PS1MAZ&6.588&0.522$_{-0.139}^{+0.177}$&11&45.50$_{-0.13}^{+0.13}$&11\\
SDSS J220226.77+150952.3&330.6115&15.1646&5.070&WISEHI&5.070&0.50$_{-0.17}^{+0.16}$&6&45.15$_{-0.18}^{+0.12}$&6\\
SDSS J221644.01+001348.1&334.1834&0.2300&5.010&DR16Q&5.010&0.23*$ _{-0.12}^{+0.11}$&14,1*,2&45.00$_{-0.40}^{+0.30}$&14\\
PSO J338.2298+29.5089&338.2298&29.5089&6.658&PSO&6.666&0.141$_{-0.083}^{+0.130}$&5,11&45.76$_{-0.38}^{+0.29}$&5\\